\documentclass[twoside,twocolumn]{article}

\usepackage[sc]{mathpazo} % Use the Palatino font
\usepackage[T1]{fontenc} % Use 8-bit encoding that has 256 glyphs
\usepackage{lmodern} % SH added this
\usepackage{microtype} % Slightly tweak font spacing for aesthetics

\usepackage[english]{babel} % Language hyphenation and typographical rules

\usepackage{graphicx}
\usepackage{color}
\usepackage{dcolumn}% Align table columns on decimal point
\usepackage{bm}% bold math
\usepackage[top=2cm,bottom=2cm,left=1.5cm,right=1.5cm,columnsep=20pt]{geometry} % Document margins
    {\large\scshape Computational Details%
    \par\medskip\normalfont\normalsize}%
    {}%
\newenvironment{acknowledgement}%       New acknowledgement environment
    {\large\scshape Acknowledgement%
    \par\medskip\normalfont\normalsize}%
    {}%
    {\large\scshape Supporting Information%
    \par\medskip\normalfont\normalsize}%
    {}%
\usepackage{multirow} % SH added for table
\usepackage{slashbox} % SH added this for table

\usepackage[utf8]{inputenc}
\usepackage[T1]{fontenc}
\usepackage{mathptmx}
\usepackage[most]{tcolorbox}

\usepackage[normal, small,labelfont=bf,up,textfont=it,up]{caption} % SH added this (180704)
\usepackage{booktabs} % Horizontal rules in tables

\usepackage{lettrine} % The lettrine is the first enlarged letter at the beginning of the text

\usepackage{enumitem} % Customized lists
\setlist[itemize]{noitemsep} % Make itemize lists more compact

\usepackage{abstract} % Allows abstract customization
\usepackage{titlesec} % Allows customization of titles
\renewcommand\thesection{\Roman{section}} % Roman numerals for the sections
\renewcommand\thesubsection{\roman{subsection}} % roman numerals for subsections
\titleformat{\section}[block]{\large\scshape\centering}{\thesection.}{1em}{} % Change the look of the section titles
\titleformat{\subsection}[block]{\large}{\thesubsection.}{1em}{} % Change the look of the section titles

\usepackage{fancyhdr} % Headers and footers
\usepackage{titling} % Customizing the title section

\usepackage{hyperref} % For hyperlinks in the PDF

\usepackage{verbatim} %multiline comment, \begin{comment} ~ \end{comment}

\usepackage{amstext} %available planetxext in equations, $\text{ abc }$

\usepackage{amsmath} 

\usepackage[numbers]{natbib}

\pretitle{\begin{center}\Large\bfseries} % Article title formatting
\posttitle{\end{center}} % Article title closing formatting

\def\bea{\begin{eqnarray}}
\def\eea{\end{eqnarray}}
\def\ben{\begin{equation}}
\def\een{\end{equation}}
\def\benu{\begin{enumerate}}
\def\enu{\end{enumerate}}

\def\bei{\begin{itemize}}
\def\eei{\end{itemize}}
\def\beit{\begin{itemize}}
\def\eit{\end{itemize}}
\def\benu{\begin{enumerate}}
\def\enu{\end{enumerate}}

\def\sss{\scriptscriptstyle\rm}
\def\br{{\bf r}}
\def\tar{^{\sss tar}}
\def\b{^{\sss b}}
\def\ext{_{\sss ext}}
\def\H{_{\sss H}}

\def\s{_{\sss S}}
\def\xc{_{\sss XC}}

\definecolor{Red}{rgb}{0.816,0.02,0}
\definecolor{Green}{rgb}{0.016,0.627,0}
\definecolor{Plum}{rgb}{0.17,0,0.45}
\definecolor{LBlue}{rgb}{0,0.34,0.45}
\definecolor{Sepia}{rgb}{0.37,0.17,0.02}
\definecolor{BurntOrange}{rgb}{0.88,0.39,0}
\definecolor{dkgreen}{rgb}{0,0.6,0.5}
\definecolor{gray}{rgb}{0.5,0.5,0.5}
\definecolor{black}{rgb}{0,0,0}
\definecolor{white}{rgb}{1,1,1}
\definecolor{grey}{rgb}{0.75,0.75,0.75}
\definecolor{mauve}{rgb}{0.58,0,0.82}
\definecolor{lightyellow}{rgb}{0.98,0.98,0.94}

\newcommand{\codebox}{\begin{tcolorbox}[colback=grey,colframe=black,arc=0pt]}
\newcommand{\codeboxend}{\end{tcolorbox}}
\newcommand{\terminalout}{\begin{tcolorbox}[colback=lightyellow,colframe=black,arc=0pt]}
\newcommand{\terminaloutend}{\end{tcolorbox}}
\newcommand{\code}[1]{{\fontfamily{qcr}\selectfont{#1}}}

\title{KS-pies: Kohn-Sham Inversion Toolkit}
\author{%
\textsc{Seungsoo Nam, Hansol Park, and Eunji Sim}\thanks{esim@yonsei.ac.kr} \\ % name
\normalsize Department of Chemistry, Yonsei University, 50 Yonsei-ro Seodaemun-gu, Seoul 03722, Korea \\ % institution
\textsc{Ryan J. McCarty} \\ % name
\normalsize Departments of Chemistry, University of California, Irvine, CA 92697, USA \\  % institution
}

\date{\today}% It is always \today, today,
\renewcommand{\maketitlehookd}{
\begin{abstract}
A Kohn-Sham (KS) inversion determines a KS potential and orbitals corresponding to a given electron density, a procedure that has applications in developing and evaluating functionals used in density functional theory.
Despite the utility of KS inversions, application of these methods among the research community is disproportionately small. We implement the KS inversion methods of Zhao-Morrison-Parr and Wu-Yang in a framework that simplifies analysis and conversion of the resulting potential in real-space.
Fully documented Python scripts integrate with PySCF, a popular electronic structure prediction software, and  Fortran alternatives are provided for computational hot spots. 
\end{abstract}
}

\begin{document}

% Print the title
\maketitle

%%-----------------------------------------------------------------------------------------------------------
%	ARTICLE CONTENTS
%%-----------------------------------------------------------------------------------------------------------

%\section{Introduction}

%\lettrine[nindent=0em,lines=3]{L} orem ipsum dolor sit amet, consectetur adipiscing elit.

%\subsection{Subsection One}
%A statement requiring citation\cite{Figueredo:2009dg}.
%\subsection{Subsection Two}

%%%%%%%%%%%%%%    OUTLINE (BEGIN)  %%%%%%%%%%%%%%%%%%
\iffalse
??
\fi
%%%%%%%%%%%%%%    OUTLINE (END) %%%%%%%%%%%%%%%%%%%

\sf
%%%%%%%%%%%%%%%%%%%%%%%%%%%%%%%%%%%%%%%%%%%%%%%%%%%%%%%%%%%%%%%%%%

\section{\label{intro} Introduction}
Future progress in improving Kohn-Sham (KS) density functional theory (DFT), one of the most popular computational techniques for materials and molecules, depends upon improved exchange-correlation (XC) functionals.\cite{B12}
KS DFT assumes that non-interacting electrons and a local, multiplicative potential (i.e. KS potential) approximate the electron density of the real system's interacting electrons.
No systematic method for improving XC functionals has been identified within the KS system. Instead, functional development draws upon a broad array of techniques, methods, and data sources for guidance on how to produce more accurate approximations. One resourceful method is the KS inversion, which produces a KS potential from a provided electron density. 
Insight from KS inversions have been used in functional development since 1996 \cite{TIH96} and has seen a recent revival in its use today.\cite{NAST18, NOL19, ZWCC19}
It has also become a useful tool for studying DFT methods, such as time-dependent DFT,\cite{BHUG13, KOS19} density-corrected DFT, \cite{NSSB20} inter-molecular interactions in partition-DFT \cite{EBCW10, NW14} and embedded-DFT,\cite{GAMM10,BA18,ZC18} and even in symmetry-adapted perturbation theory.\cite{BJ19}

A KS inversion can construct an exact potential from an exact electron density as defined by the one-to-one density-to-potential mapping stated in the Hohenberg-Kohn theorem.\cite{HK64}
However, in practice, the finite number of localized basis sets needed to expand KS orbitals destroys the advantageous one-to-one mapping,\cite{HIGB01, SSD06,JW18} and the problem becomes ill-posed.\cite{JW18} Although an exact KS inversion is no longer possible, several approximate methods\cite{AP84, G92, WP93, VB94, ZMP94, WY03, RS12, ZC18, FAB18, KZG19, CLG20} have been proposed.

Despite the developed theory and applications of KS inversions, publicly available software for routinely performing these calculations is not common.\cite{Serenity}
In this work, we introduce KS-pies, an open source implementation of the most frequently cited KS inversion methods, Zhao-Morrison-Parr\cite{ZMP94} (ZMP) and Wu-Yang\cite{WY03} (WY).
Our WY implementation supports user-defined potential basis sets and Hamiltonians.
This software includes a utility module that helps simplify inversion calculations by providing input file conversion function and and real-space evaluation of inversion potentials. The software is distributed under an open source Apache 2.0 licence and developer community is hosted though GitHub.

Our Python implementation utilizes NumPy,\cite{NumPy} SciPy libraries,\cite{SciPy} and features from PySCF\cite{PySCF18} that are familiar to the DFT community. A section of Python code can alternatively call \code{kspies\_fort}, a compiled Fortran version that decreases computational cost.

In this paper we include a theoretical summary, details on implementation, validation, and an appendix with a discussion based user-guide connecting technical features with examples that highlight the simplicity of incorporating KS inversions into modern functional development workflows. Emphasis on the theoretical background seeks to provide a simplified and practical entry point for theorists unfamiliar with the methodology.

\section{Background}

\subsection{Kohn-Sham Density Functional Theory and its Inverse}

The conventional (forward) KS procedure solves a single particle equation 
\begin{equation}
\{ -\frac{1}{2}\nabla^2 +v\s[n](\br) \} \psi_i(\br)=\varepsilon_i\psi_i(\br),
\label{eq:KS}
\end{equation}
where $\psi_i$ and $\varepsilon_i$ are $i$-th KS orbitals and orbital energies, $v\s[n](\br)$ is the KS potential, and $n$ is the electron density. Atomic units are used throughout unless specified.
Typically $v\s[n](\br)$ is written as
\begin{equation}
v\s[n](\br) = v_{\sss ext}(\br) + v\H[n](\br) +v\xc[n](\br),
\end{equation}
where $v_{\sss ext}(\br)$ is the external potential, $v\H[n](\br)$ is the Hartree potential
\begin{equation}
v\H[n](\br)=\int{ \frac{n(\br')}{\|\br - \br'\|} d\br'},
\end{equation}
and $v\xc[n](\br)$ is an XC potential, which is approximate in practice.
The electron density is determined by the sum of the occupied orbital densities as
\begin{equation}
n(\br) =\sum_i^{N_{\sss occ}} | \psi_i (\br)|^2.
\label{eq:den}
\end{equation}
As a funcitonal of the density, $v\s[n](\br)$ is used in Eq.~\ref{eq:KS} in a self-consistent field (SCF) procedure, until a converged electron density is determined.

KS inversions operate in reverse, using a given density (often refered as target density, $n\tar(\br)$) to determine $v\s$.
Once determined, $v\s$ and Eq.~\ref{eq:KS} produce the KS orbitals and associated eigenvalues.
In principle, one-to-one density-to-potential mapping stated in the Hohenberg-Kohn theorem\cite{HK64} guarantees that KS inversion can construct an exact KS potential (up to a constant) from an exact electron density.
However, in practice, the finite number of localized basis sets needed to expand KS orbitals destroys the advantageous one-to-one mapping,\cite{HIGB01, SSD06,JW18} and the problem becomes ill-posed.\cite{JW18} Although an exact KS inversion is no longer possible, several approximate methods\cite{AP84, G92, WP93, VB94, ZMP94, WY03, RS12, ZC18, FAB18, KZG19, CLG20} have been proposed.
KS inversions have been developed for use with input orbitals \cite{GRS13, ZC18} or wavefunctions,\cite{RKS15} but we focus exclusively on density-based methods,\cite{AP84, G92, WP93, VB94, ZMP94, WY03, RS12, FAB18, KZG19, CLG20} including ZMP\cite{ZMP94} and WY,\cite{WY03} which will be explained in the following.

\subsection{Zhao-Morrison-Parr}
The ZMP KS inversion method\cite{ZMP94} minimizes an objective self-repulsion functional
\begin{equation}
C[n^{\lambda}] = \iint \frac{[n^{\lambda}(\br) - n\tar(\br) ] [ n^\lambda(\br') -n\tar(\br') ]}{\|\br-\br'\|} d\br d\br',
\label{eq:self_rep}
\end{equation}
by solving a KS-like equation self-consistently under a given Lagrange multiplier $\lambda$
\begin{equation}
\{ -\frac{1}{2}\nabla^2 +v\s[n\tar,n^\lambda](\br) \} \psi_i^\lambda(\br)=\varepsilon_i^\lambda\psi_i^\lambda(\br).
\label{eq:ZMP}
\end{equation}
A $\lambda$-dependent KS-potential
\begin{equation}
\begin{split}
v\s[n\tar,n^\lambda](\br) =& v\ext(\br)+v\H[n\tar](\br) \\
& + v_g[n\tar](\br)+v_C^\lambda[n\tar,n^\lambda](\br),
\end{split}
\label{eq:ZMPvs}
\end{equation}
includes a guiding potential $v_g(\br)$ and correction potential 
\begin{equation}
v_C^\lambda[n\tar,n^\lambda](\br)=\lambda \int \frac{n^\lambda(\br') - n\tar(\br')}{\| \br - \br' \|} d\br' .
\label{eq:ZMPvC}
\end{equation}
In principle, as $\lambda \rightarrow \infty$, $C[n^\lambda] \rightarrow 0$ and $n^\lambda \rightarrow n\tar$.
Only $v_C^\lambda$ depends on $\lambda$ in Eq. \ref{eq:ZMPvs} and accommodates all necessary potential modification. If provided, additional potential terms $v\ext(\br), v\H(\br)$, and $v_g(\br)$ accelerate the convergence of $v\s$ with respect to $\lambda$. 
In practice, Eq.~\ref{eq:ZMP} is solved self-consistently using a given $\lambda$ value. The obtained orbitals are used as an initial guess for following calculations at larger $\lambda$ values, a process repeated until $\lambda$ become large enough.

The guiding potential $v_g(\br)$ mimics the XC potential, for which a variety of potentials can be used.
Typically $v_g(\br)$ is initially formulated to mimic the asymptotic decay of XC potential, $-(1/N)v\H(\br)$, where $N$ is the number of electrons.
We refer $-(1/N)v\H(\br)$ as FAXC, the non-Hartree portion of the Fermi-Amaldi potential.\cite{FA34} In principle, any potential can be used for $v_g(\br)$ when the asymptotic decay of XC potential is not important.\cite{NSSB20}

For open-shell systems, ZMP is used as a spin-unrestricted formalism.\cite{THG97}
The correction potential is spin-dependent, and Eq.~\ref{eq:ZMPvC} is rewritten for $\alpha$ or $\beta$ spin as
\begin{equation}
v_{C,\sigma}^\lambda[n\tar_\sigma,n^\lambda_\sigma](\br)=2\lambda \int \frac{n^\lambda_\sigma(\br') - n\tar_\sigma(\br')}{\| \br - \br' \|} d\br',
\label{eq:ZMPvC_spin}
\end{equation}
where $\sigma$ denotes the spin index. 
The factor of 2 is required for consistent results in closed-shell systems for restricted and unrestricted schemes.
The guiding potential is spin-dependent for standard DFT XC potentials, but not for FAXC.

\subsection{Wu-Yang}

The WY approach \cite{WY03} maximizes an objective functional, 
\begin{equation}
\begin{split}
W\s[\{ b_t \}]= &\sum_i^{N/2} \int |\nabla\psi\b_i(\br)|^2 d\br \\
& + \int v\s\b[n\tar](\br) \{ n\b(\br)-n\tar(\br) \} d\br,
\end{split}
\label{eq:WY}
\end{equation}
where $v\s\b(\br)$ is a KS-potential similar to Eq.~\ref{eq:ZMPvs} in ZMP, except with $v_C(\br)$ represented as a linear combination of potential basis functions $g_t(\br)$, given as
\begin{equation}
v_C\b(\br) = \sum_t b_t g_t(\br),
\label{eq:WY_corr}
\end{equation}
making $v\s\b(\br)$ exclusively a functional of the target density.
KS orbitals $\psi\b_i(\br)$ are determined by solving a KS-like equation,
\begin{equation}
\{ -\frac{1}{2}\nabla^2 + v\s\b[n\tar](\br) \} \psi_i\b (\br) = \varepsilon_i\b \psi_i\b (\br),
\label{eq:WY_KS}
\end{equation} 
that does not require SCF procedure.
The objective functional $W\s$ is maximized by adjusting $\{ b_t \}$.
The gradient and Hessian of $W\s$ with respect to $\{ b_t \}$ is given in an analytical form
\begin{equation}
\frac{ \partial W\s }{ \partial b_t } = \int \{n\b(\br) - n\tar(\br) \} g_t(\br) d\br,
\label{eq:WY_grad}
\end{equation}
\begin{equation}
\frac{ \partial^2 W\s }{ \partial b_t \partial b_u} 
= \sum_i^{N_{\sss occ}} \sum_a^{N_{\sss vir}}
\frac{
\langle \psi_a\b | g_t | \psi_i\b \rangle
\langle \psi_a\b | g_u | \psi_i\b \rangle
}{\varepsilon_i\b - \varepsilon_a\b},
\label{eq:WY_hess}
\end{equation}
where $N_{\sss occ}$ ($N_{\sss vir}$) denotes the number of occupied (virtual) orbitals, simplifying maximization of Eq.~\ref{eq:WY}.

In a spin-unrestricted formalism, Eq.~\ref{eq:WY} becomes
\begin{equation}
\begin{split}
W\s[\{ b_t \}]&= \frac{1}{2}\sum_\sigma \sum_i^{N_\sigma} \int |\nabla\psi\b_{i,\sigma}(\br)|^2 d\br \\
& + \int \{ v\ext(\br) + v\H[n\tar](\br) \} \{ n\b(\br)-n\tar(\br) \} d\br \\
& + \sum_\sigma \int \{ v_{g,\sigma}[n\tar,n\tar_\sigma](\br)+v_{C,\sigma}\b(\br) \}  \\
&~~~~~~~~~~~~~~~~~~\times \{ n\b_\sigma(\br)-n\tar_\sigma(\br) \} d\br.
\end{split}
\end{equation}
Spin dependence of XC potentials is reflected in the correction ($v_C$) potential due to its spin dependence on $\{ b_t \} $ and standard DFT XC guiding ($v_g$) potentials.

Highly oscillatory XC potentials are a well-known drawback of the WY method that often arise when using a large potential basis.\cite{HBY07}
One solution to this problem is regularization with the objective functional

\begin{equation}
\overline{W}\s^\eta(\{ b_t\})=W\s(\{ b_t\})+\eta \| \nabla v_C\b(\br) \|^2,
\label{eq:WY_reg}
\end{equation}
where $\eta$ is a regularization strength hyperparamater, and the smoothness of the correction potential is measured with
\begin{equation}
\| \nabla v_C\b(\br) \|^2 = \int v_C\b(\br) \nabla^2 v_C\b (\br) d\br.
\label{eq:Reg}
\end{equation}
An optimal value of $\eta$ must be selected. If $\eta$ is too small, the potential may contain severe oscillations, but if $\eta$ is too large, the optimized potential becomes too smooth and misses physically important features.
Heaton-Burgess and Yang\cite{HY08} describes how this optimal selection can be made.

\section{Implementation}

\subsection{Architecture and General workflow of KS-pies}

KS-pies consists of three sub-modules; \code{zmp} and \code{wy}, which are the modules for KS inversion, and \code{util} module that supports additional utility functions to help prepare inputs to \code{zmp} and \code{wy} or analyze inversion results.
To start inversion calculation (ZMP or WY) with KS-pies, two inputs are required: \code{Mole} object and density matrix of the target density.
\code{Mole} object defines standard details needed for quantum chemical calculations, such as atomic coordinates, number of electrons, and basis sets.
This is exactly the \code{Mole} object defined in PySCF, which can be defined with only a few lines of code or loaded from a common geometry file such as xyz.
The second input, density matrix, can be generated with PySCF calculations or loaded from external file format such as \code{molden}.\cite{Molden}
To extend accessibility to KS-pies and PySCF, \code{util} sub-module in KS-pies also supports loading \code{wfn} file format, which is supported for variety of quantum chemistry programs, such as Gaussian,\cite{g16} ORCA,\cite{ORCA} GAMESS,\cite{GAMESS} and Molpro.\cite{Molpro}
Target densities generated with quantum chemistry programs that support \code{molden} or \code{wfn} formats can be used as inputs for KS-pies.

Performing an inversion calculation with KS-pies is very similar to running KS-DFT calculation with PySCF.
User can specify algorithm-dependent options for running ZMP and WY calculations.
All inversion calculations are defined as a class object, and KS-pies manages data using Python instance variables, storing the majority of results in memory.
Data from one instance can be used as an initial guess for subsequent calculations or analysis within KS-pies or PySCF.
\code{util} module supports additional utility functions to help prepare inputs to \code{zmp} and \code{wy} and analyze inversion results.

KS-pies makes use of analytical functions within the PySCF integral library \cite{libcint} to convert Eq.~\ref{eq:ZMP} and ~\ref{eq:WY_KS} to solvable matrix equations. 
The Hartree and guiding potentials are constructed from the target density using a matrix representation by default.
KS-pies then uses a method-specific procedure to optimize the correction potential.
For ZMP, a self-consistent calculation with Eq.~\ref{eq:ZMP} and a user provided $\lambda$ is performed. 
For WY, the obtained gradient (Eq.~\ref{eq:WY_grad}) and Hessian (Eq.~\ref{eq:WY_hess}), $\{b_t\}$ are adjusted to maximize Eq.~\ref{eq:WY} using \code{SciPy} optimizer.

WY can accommodate non-Gaussian potential basis sets. Although a Gaussian function is typically used to expand potentials due to its integration efficiency, Eq.~\ref{eq:WY_corr} is not limited to Gaussian type basis set. 
KS-pies accommodates this uncommon feature and can handle user-defined potential basis set.
When arbitrary user-defined potential basis sets are encountered, a numerical integration of the three-center overlap integral used in Eq.~\ref{eq:WY} is calculated with
\begin{equation}
S_{ijt} = \int \phi_i(\br) \phi_j(\br) g_t(\br) d \br,
\label{eq:Sijt}
\end{equation}
where $\phi_i(\br)$ is $i$-th orbital basis function. Numerical integration of Eq.~\ref{eq:Sijt} adds a substantial computational overhead at the beginning of the calculation.

Several features reduce the computational cost of determining the correction potential. For ZMP, recalculating the Hartree potential at each self-consistent iteration can be accelerated using a density fitting procedure.
Density fitting results in minor differences in the inversion, while greatly reducing the computational cost.
For WY, Eq.~\ref{eq:WY_grad} and Eq.~\ref{eq:WY_hess} are calculated using a Fortran module \code{kspies\_fort} to provide a substantial time savings over a Python implementation.
Compiled binaries and the source code are provided.

\subsection{Real-Space Potentials}
The KS potential in real-space provides valuable insight beyond a target density calculation.
For example, a major motivation for performing a KS inversion is the visualization of the exact KS potential that can show the non-intuitive step structures present in some potentials.\cite{VGB95, KPS16}
However, the use of finite atom-centered Gaussian basis-sets prevents the XC potentials produced by ZMP and WY from being directly converted into real-space representation. 
For a given system, the basis functions $\{ \phi_i(\br)\}$, real-space function $v(\br)$ and its matrix representation $V_{ij}$ have the relation
\begin{subequations}
\begin{eqnarray}
v(\br) \rightarrow V_{ij} = \int \phi_i(\br) v(\br) \phi_j(\br) d\br, \\
V_{ij} \rightarrow v(\br) = \sum_{ij}\phi_i(\br) V_{ij} \phi_j(\br).
\end{eqnarray}
\label{eq:conversion}
\end{subequations}
However, Eq.~\ref{eq:conversion}b is only exact under the basis set limit, and in practice, would result in large errors in the real-space function $v(\br)$.
Therefore, a method other than Eq.~\ref{eq:conversion}b is required to obtain real-space values of $v(\br)$.
This is necessary when the guiding or correction potential needs to be evaluated in real-space.
The method of Franchini et al.\cite{FPLV14} for converting Hartree potentials can be applied to the FAXC guiding potential and the ZMP correlation potential.
A DFT XC potential on real-space is necessary when ZMP or WY utilizes a DFT XC guiding potential.

KS-pies evaluates real-space Hartree potentials following Franchini et al.\cite{FPLV14} In this approach, the density is decomposed into one-center (i.e. atomic) contributions, and into different angular contributions as 
\begin{equation}
n(\br) = \sum _i^{N_{nuc}} n_i(\br) \approx  \sum _i^{N_{nuc}}\sum_l^{l_{\sss max}}\sum_m Z_{lm}(\theta_i, \varphi_i) s_{lm}^i(r_i),
\label{eq:vH1}
\end{equation}
where $Z$ represents real spherical harmonics and $s$ is a cubic spline interpolation of radial density at the $i$-th atoms radial grid.
The summation inside Eq.~\ref{eq:vH1} can be used to calculate the Hartree potential for each angular contribution of $i$-th atom with analytical form
\begin{equation}
\begin{split}
v_{{\sss H},lm}(\br_i) &= \frac{4\pi}{2l+1} Z_{lm}(\theta_i,\varphi_i) \\
&\times \Big( \frac{1}{r_i^{l+1}} \int_0^{r_i} r'^{l+2}s_{lm}^i(r') dr' \\
&+r_i^l \int_{r_i}^{\infty} \frac{ s_{lm}^i (r')}{{r'}^{l-1}} dr' \Big).
\end{split}
\label{eq:vH2}
\end{equation}
This allows conversion of a density to a real-space Hartree potential directly, without going through matrix representation, and is implemented in \code{kspies.util.eval\_vH}.

The \code{kspies.util.eval\_vxc} function evaluates DFT XC potential on user defined grid points. Our implementation is based on the numerical differentiation.
For local density approximation functionals (LDA) we use, 
\begin{equation}
v\xc^{\sss LDA} (\br) = \frac{\delta E\xc}{\delta n} = \frac{d\epsilon\xc^{\sss LDA}}{dn(\br)},
\end{equation}
where $\epsilon\xc^{\sss LDA}$ is the XC density of LDA, and can be directly obtained from \code{pyscf.dft.libxc.eval\_xc}.
For generalized gradient approximation (GGA) functionals we use, 
\begin{equation}
v\xc^{\sss GGA} (\br) = v_n -2\{ \nabla n \cdot \nabla v_\gamma + v_\gamma \nabla^2 n \},
\label{eq:vgga}
\end{equation}
where $v_n = \partial \varepsilon\xc^{\sss GGA}/ \partial n$, $v_\gamma = \partial \varepsilon\xc^{\sss GGA}/\partial \gamma$, and $\gamma = \nabla n \cdot \nabla n $. 
Although $v_n$ and $v_\gamma$ are obtainable from \code{pyscf.dft.libxc.eval\_xc}, $\nabla v_\gamma $ should be evaluated using a numerical derivative of $v_\gamma$.
For spin-polarized densities, Eq.~\ref{eq:vgga} for $\alpha$ spin becomes
\begin{equation}
\begin{split}
v_{{\sss XC},\alpha}^{\sss GGA} (\br) 
= v_{n_\alpha} - 2 ( \nabla n_\alpha \cdot \nabla v_{\gamma_{\alpha \alpha}} + v_{\gamma_{\alpha \alpha}}  \nabla^2 n_\alpha ) \\
- ( \nabla v_{\gamma_{\alpha \beta}} \cdot \nabla n_\beta + v_{\gamma_{\alpha \beta}} \nabla^2 n_\beta ),
\end{split} 
\label{eq:vgga_s}
\end{equation}
and the formulation for $\beta$ spin requires a trivial swapping of respective spins.
Eq.~\ref{eq:vgga} and \ref{eq:vgga_s} are implemented in \code{kspies.util.eval\_vxc}.
Section \ref{Utility_section} Utility and associated figures provide an example of obtaining real-space representation using these approaches.

\section{Validation and performance} 

Valid implementation is confirmed with accurate KS inversions of densities obtained from HF or correlated wavefunction methods, in both restricted and unrestricted schemes. Run time benchmarks are reported as wall time and were performed an Intel(R) Xeon(R) Gold 6142 CPU using 8 processors at 2.6 GHz. Calls to PySCF and \code{kspies\_fort} can take advantage of  parallelization with OpenMP.
In our KS inversion examples below, benzene used the most memory, approximately 1.2 GB. The PySCF CCSD Calculation of molecular O$_2$ used substantially more memory, however, the inversion using KS-pies required less than that of benzene.
Restricted and unrestricted inversion benchmarks are included for convenience within the software repository. 

\subsection{Restricted Inversion}

\begin{figure}[!ht]
\centering
\includegraphics[width=1\columnwidth]{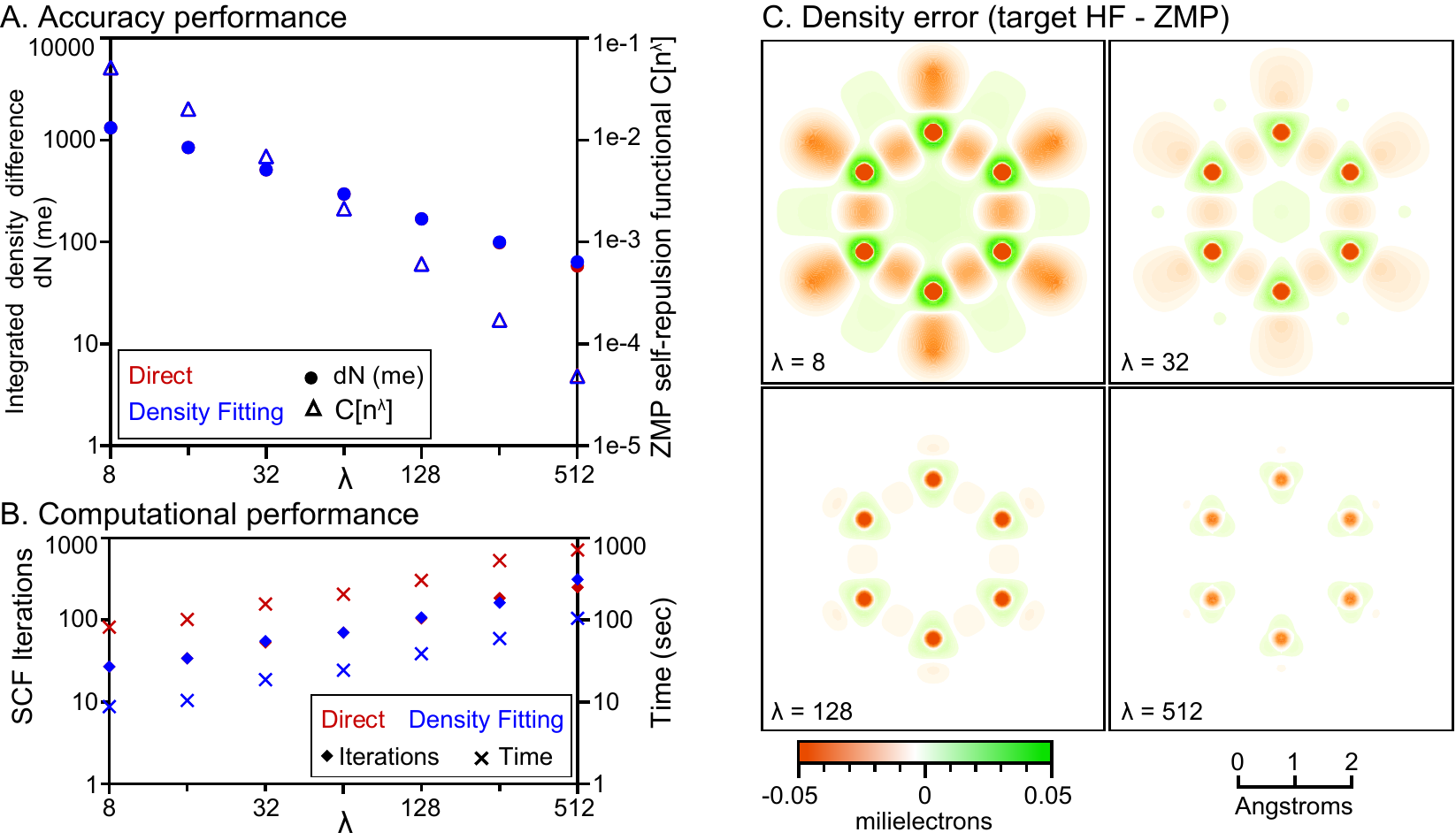}
\caption{A comparison of ZMP results with and without density fitting. For each $\lambda$ value, \code{level\_shift} was set to $0.1 \times \lambda$. (a) Accuracy performance evaluated by integrated density difference in milielectron (dN) and  minimized self-repulsion functional value (Eq.~\ref{eq:self_rep}). (b) Computational performance in terms of SCF iterations and time needed for convergence. 
(c) Density difference between target HF density and ZMP densities for increasing $\lambda$.
}
\label{fig:bz}
\end{figure}

Restricted inversion performance was validated using ZMP and WY on benzene ($R_{\sss CC}=1.3936$ \AA  and $R_{\sss CH}=1.0852$ \AA). The target density was generated with HF/cc-pVTZ. Resulting inversion potentials from ZMP and WY can be used to accurately reproduce the target density. Furthermore, ZMP, qualitatively indicates that $C$ (Eq.~\ref{eq:ZMPvC}) is approaching $0$ as $\lambda$ increases. 

\code{kspies.zmp.RZMP} (restricted ZMP) was used with a FAXC guiding potential and tested with and without density fitting. Using the self-repulsion functional value $C$ from Eq.~\ref{eq:self_rep} and the integrated density difference $dN = \int | n^\lambda(\br)-n\tar(\br)| d\br$, we confirmed an expected decrease in $C$ as $\lambda$ increased. This is plotted in Figure~\ref{fig:bz}a.
As $\lambda$ increases, $dN$ also decreases. Our implementation shows negligible difference in accuracy with and without density fitting.

Computationally, the density fitting method decreases the cost of ZMP.
Figure~\ref{fig:bz}b plots computational performance in terms of SCF iterations and run time per $\lambda$; highlighting negligible difference in the number of SCF cycles.
Per SCF iteration, ZMP takes approximately 0.25 seconds with density fitting and 3 seconds without, requiring around 700 iterations to reach convergence on small molecules.

WY performance was evaluated using the same target density as above and cc-pVTZ for the potential basis set.
Optimization was complete after 8 iterations, taking less than 1 second on default settings.
The maximum gradient element was $3 \times 10^{-8}$ with  $dN=170.8$~me.
The $dN$ agrees with the values from ZMP at $\lambda$=128, indicating an accurately reproduced target density.
Implementation verification is also incidentally discussed in section \ref{WY_userdefined_section} where WY reproduces a target density from a user-defined harmonic potential Hamiltonian.

\subsection{Unrestricted Inversion}
To validate the unrestricted calculations of ZMP and WY, we use a coupled-cluster singles-and-doubles (CCSD) target density of molecular oxygen ($R_{\sss OO}=1.208$ \AA) obtained with UHF-UCCSD/cc-pVQZ. We used FAXC guiding potential for both cases.

Benchmark values for ZMP with $\lambda$=2048 are $C$=1.10$\times$10$^{-6}$ and $dN$=5.75~me.
A small and decreasing $dN$ as $\lambda$ is increased confirms the unrestricted ZMP implementation.
As a secondary validation benchmark, identical results are produced from spin-restricted and unrestricted ZMP calculation on closed-shell benzene.

Benchmark values for WY with a cc-pVQZ potential basis set converges after 5 optimization steps, taking 0.07 seconds, with a maximum gradient element of 3$\times$10$^{-8}$, and dN=36.3~me, which agrees with ZMP $dN$ at $\lambda$=128, verifying the WY implementation.

\subsection{Inversion of User-Defined Systems}

\begin{figure}[!ht]
\centering
\includegraphics[width=\columnwidth]{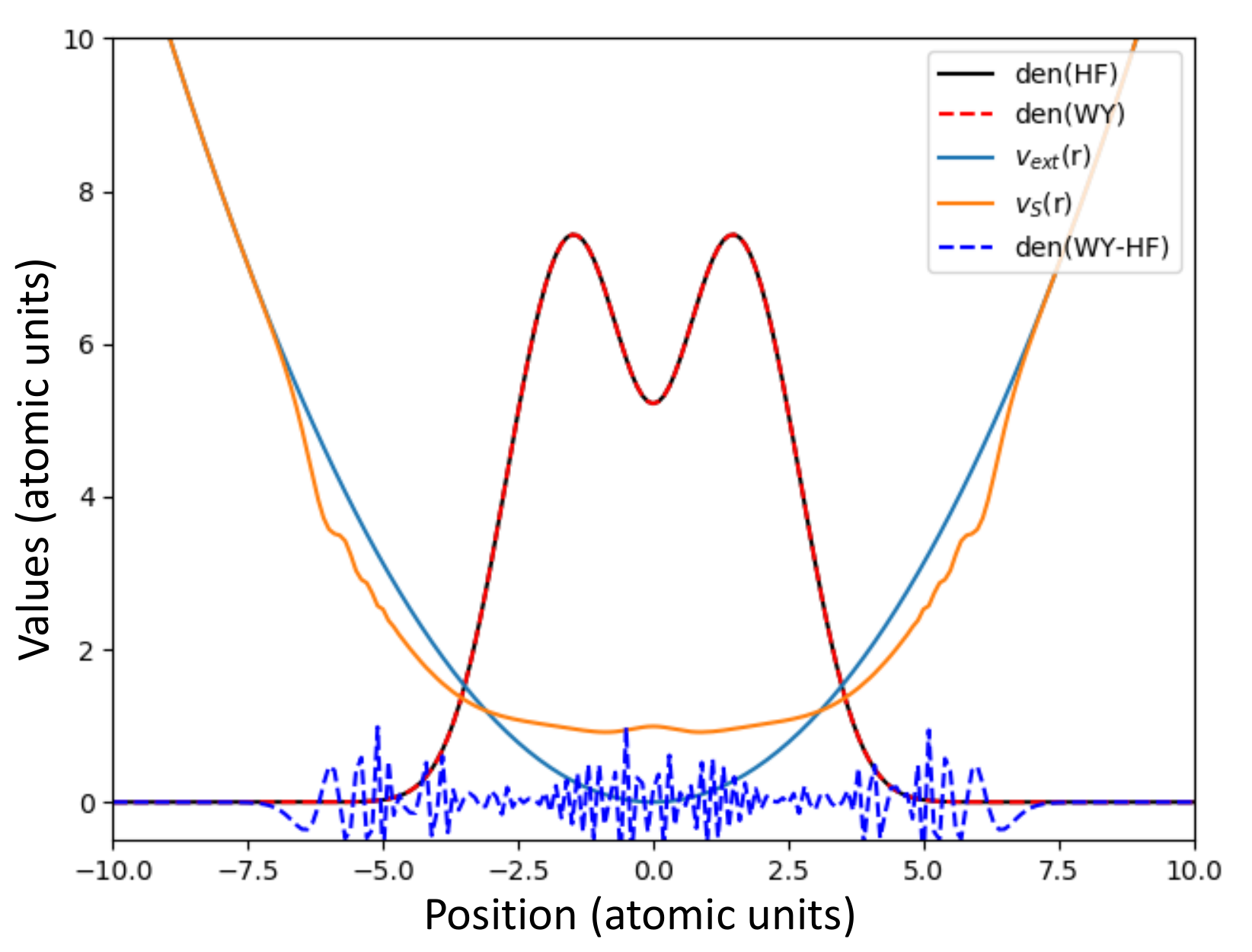}
\caption{
Finite-difference HF target density (black) generated from harmonic external potential (blue) and inverted density (red dashed) and KS potential (orange) generated with WY. 
For visibility, densities are increased vertically 10$\times$ and  density differences are increased 10$^6\times$.
}
\label{fig:Harm}
\end{figure}

WY inversions can be calculated on user-defined systems using \code{kspies.wy} and a properly defined, user supplied, Hamiltonian component. 
Figure~\ref{fig:Harm} shows a calculated potential from a user-defined Hamiltonian of a harmonic potential ($v_{\sss ext}(x)=1/8 x^2$,blue solid curve in Figure~\ref{fig:Harm}) with four electrons. 
Finite-difference HF with soft electron-electron repulsion $w(x_1,x_2)=1/\sqrt{(x_2-x_1)^2+0.5^2}$ was used to generate the target density within the domain $x \in [-10, 10]$ and a grid spacing of 0.02.
The WY calculation was then performed using the finite-difference method by providing a finite-difference Hamiltonian and using a grid to expand the potential.

Section \ref{WY_userdefined_section} and the online KS-pies documentation expand upon the present example with information on how to obtain necessary inputs for user-defined Hamiltonians.

The resulting WY KS potential for the user-defined harmonic potential is displayed in orange in Figure ~\ref{fig:Harm}.
The KS potential includes small oscillations near $x=\pm$7 due to a known issue of WY\cite{JW18} arising from the small electron density in these regions.
The oscillations are not a fault of our implementation.
The resulting potential, and its oscillations are not unique.\cite{J11} In the general use of WY, these oscillations can be influenced by conditions, such as the optimization scheme, convergence criteria, or selection of a basis set.
Nevertheless, the agreement of the density as represented by the red and black curves in Figure ~\ref{fig:Harm} highlights the accuracy of WY theoretically, as well as our successful implementation.

\section{Usage}
To encourage the use of KS inversions in functional development, \code{kspies} requires only a few inputs. In appendix \ref{appendix:examples} we provide several examples that highlight the relative simplicity of routine KS inversions with and without real space conversions.
The ZMP section details basic ZMP use, and the WY section details basic use as well as an example using regularization and another using a user defined harmonic potential.
Utility module examples demonstrate importing target densities from other software and analysis tools for evaluating and visualizing resulting KS potentials.

The package can be easily installed from PyPI using \code{pip install kspies}, and full documentation, examples from the appendix, and test scripts are available online. 
Frequent users can compile a provide Fortran subroutine that improves the speed of WY calculations.

\section{Conclusion}

KS-pies presents an open source, publicly available code for performing KS inversions of electron densities into KS potentials. Since every KS inversion method is approximate, we implemented the two most cited inversion methods, ZMP and WY. Our software integrates with PySCF, an environment familiar to the theoretical development community. Our framework provides a starting point for the implementation of future KS inversion methods. This publication presents the theoretical context and examples that highlight the simplicity of running KS inversions. Incorporating KS inversion methods to determine real-space potentials should be beneficial for XC functional development and testing.

With two implemented methods, users are able to choose and compare results, leveraging advantages of each method. ZMP requires many SCF iterations and is computationally intensive relative to WY, but the result of inversion can be systematically improved by increasing $\lambda$. Alternatively, WY can perform inversions on user-defined Hamiltonians and is computationally efficient. 
Potentials produced by both methods can be converted into real space representations using our software.

\section{Data availability}
The KS-pies code is openly available on GitHub (https://github.com/ssnam92/KSPies) and can be referenced via  https://doi.org/10.25351/V3.KS-PIES.2020

\vspace{5mm}
\begin{acknowledgement}
We thank Kieron Burke for thoughtful manuscript feedback. S.N., H.P., and E.S. are thankful for support from the National Research Foundation of Korea (NRF-2020R1A2C2007468 and NRF-2020R1A4A1017737).
R.J.M is thankful for support from the University of California President's Postdoctoral Fellowship and the National Science Foundation (CHE 1856165).
\end{acknowledgement}

\appendix

\section{Basic KS-pies Usage and Examples}
\label{appendix:examples}
\subsection{Zhao-Morrison-Parr method}

\begin{figure}[!ht]
\codebox
\begin{lstlisting}
from pyscf import gto, scf
import kspies

mol_1 = gto.M(atom='Ne',basis='aug-cc-pVTZ')
mf = scf.RHF(mol_1).run()
P_tar_1 = mf.make_rdm1()

zmp_a = kspies.zmp.RZMP(mol_1, P_tar_1)
zmp_a.zscf(8)
\end{lstlisting}
\codeboxend
\terminalout
{\fontfamily{qcr}\selectfont\footnotesize
\noindent
converged SCF energy = -128.5332728252
lambda=    8.00 niter:   11 gap=  0.6484575 dN=  155.54 C= 3.99e-03
}
\terminaloutend
\caption{Example inputs for calculating a ZMP KS inversion (top) and the terminal outputs (bottom). PySCF is used to generate a Ne density, which is used in a ZMP KS inversion with $\lambda=8$}
\label{fig:ZMPexample1}
\end{figure}

ZMP calculations proceed by instantiating a \code{kspies.zmp.RZMP} object and then calling \code{zscf(l)}, which generates a KS potential with the user specified \code{l}, ($\lambda$), value.
The instance requires a Mole() object that defines the basis set, and an atomic orbital representation of the target density.
Calculation results are printed to the terminal, and a matrix representation form of the potential is stored as instance attributes.
This can be converted into real-space representation with the later discussed \code{kspies.util} module.

Figure~\ref{fig:ZMPexample1} demonstrates generating a Ne HF density using PySCF, generating a KS potential with \code{kspies.zmp}, and the run outputs printed to the terminal. 
Creating a \code{RZMP} instance requires a Mole() object (\code{mol\_1}) and density matrix of the target density (\code{P\_tar\_1}).
In the output, \code{niter} is the number of SCF iterations needed to reach convergence, \code{gap} is the HOMO-LUMO gap in atomic unit, \code{dN} is the integrated density difference in millielectrons, and \code{C} is the minimized value from Eq.~\ref{eq:self_rep}.
In unrestricted ZMP, resulting \code{C} values account for contributions from both spins, $2(C_\alpha + C_\beta)$.

\begin{figure}[!ht]
\codebox
\begin{lstlisting}
zmp_b = kspies.zmp.RZMP(mol_1, P_tar_1)
zmp_b.diis_space = 30
zmp_b.max_cycle = 200
zmp_b.guide = 'pbe'
zmp_b.conv_tol_dm = 1e-10
zmp_b.conv_tol_diis = 1e-7
for l in [ 8, 32, 128, 512]:
  zmp_b.level_shift = l*0.1
  zmp_b.zscf(l)

zmp_b.level_shift = 0.
zmp_b.zscf(512)
P = zmp_b.make_rdm1()
print('Expectation value:',mf.energy_tot(P))
print(zmp_b.mo_energy[2:6])
print('Converged?', zmp_b.converged, zmp_b.l)
\end{lstlisting}
\codeboxend
\terminalout
{\fontfamily{qcr}\selectfont\footnotesize
lambda=    8.00 niter:   12 gap=  0.6435071 dN=   83.32 C= 4.89e-04\\
lambda=   32.00 niter:   18 gap=  0.6719725 dN=   32.53 C= 7.77e-05\\
lambda=  128.00 niter:   26 gap=  0.6896898 dN=    9.89 C= 7.56e-06\\
lambda=  512.00 niter:   33 gap=  0.7009582 dN=    3.72 C= 7.25e-07\\
lambda=  512.00 niter:    9 gap=  0.7009582 dN=    3.72 C= 7.25e-07\\
Expectation value: -128.5330990412423\\
$[$-0.63510454 -0.63510454 -0.63510454  0.0658537 $]$\\
Converged? True 512
}
\terminaloutend
\caption{Example inputs for ZMP with DIIS settings at multiple $\lambda$ values using the \code{mol\_1} and \code{P\_tar\_1} from Figure~\ref{fig:ZMPexample1} (top). Terminal outputs reporting ZMP potential results at the specified $\lambda$ values and determined quantities form the final run (bottom).
}
\label{fig:ZMPexample2}
\end{figure}

Additional options in ZMP include:
\renewcommand{\labelenumi}{(\roman{enumi})}
\begin{enumerate}
\item \code{diis\_space} (int): 
DIIS space size. Default is \code{40}.
\item \code{max\_cycle} (int): 
Maximum SCF iterations. Default is \code{400}.
\item \code{guide} (character):
Guiding potential. \code{None} sets $v_g(\br)=-v_H(\br)$ (i.e. only the external and correction potential cover $v_s$, see Eq.~\ref{eq:ZMPvs}), \code{'faxc'} sets $v_g(\br)=-v\H(\br)/N$, or any DFT XC functional defined in PySCF can be specified at any additive value. For example, setting \code{'b3lyp-0.2*hf+0.2*faxc'} will remove the exact exchange portion from \code{b3lyp} potential and add replace it with the \code{faxc} guiding potential.
\item \code{level\_shift} (float):
Amount of level shift used during SCF cycles. 
Default is \code{0.2}. 
\item \code{with\_df} (boolean):
Use of density fitting. Default is \code{False}.
\item \code{conv\_tol\_dm} (float):
Density matrix convergence criteria. 
Default is \code{1e-7}.
\item \code{conv\_tol\_diis} (float):
Convergence criteria of DIIS error.
Default is \code{1e-4}.
\end{enumerate}

Results from \code{zscf(l)} are stored as the following class attributes:
\renewcommand{\labelenumi}{(\roman{enumi})}
\begin{enumerate}
\item \code{converged} (boolean):
 If convergence criteria was met during SCF iterations
\item \code{dm} (array): 
Density matrix
\item \code{mo\_coeff} (array): 
Molecular orbital coefficient
\item \code{mo\_energy} (array):
Molecular orbital energy
\item \code{mo\_occ} (array):
Orbital occupation numbers
\end{enumerate}

For \code{kspies.zmp.UZMP}, alpha and beta values for \code{dm}, \code{mo\_coeff}, \code{mo\_energy}, and \code{mo\_occ} are stored as an ordered pair tuple. 
The instance only stores one set of results; previous values will be overwritten and only the most recent calculations results are accessible.

On the first \code{zscf()} SCF cycle the target density is used as the initial density matrix guess, as demonstrated Figure~\ref{fig:ZMPexample2} with \code{P\_tar\_1}.
The guiding potential is constructed from user specifications during the first call to the function. Subsequent changes to the user specifications for the guiding potential after the first call are ignored.

The direct inversion of the iterative subspace (DIIS) procedure \cite{DIIS80} for convergence stabilization implemented in ks-pies is independent of the PySCF DIIS options. 
We recommend setting \code{diis\_space} $\leq 40$ to avoid a matrix singularity during DIIS extrapolation.
DIIS convergence stabilization is not used when \code{diis\_space} $\le$ 1. ZMP often fails to converge without DIIS, and we recommend using this feature, even for small $\lambda$.

When specified, virtual orbital energies are increased by \code{level\_shift}, which can aid convergence.
In general, setting \code{level\_shift = 0.1}$\times \lambda$ is likely to be sufficient for most systems. 
\code{level\_shift} is inactive when set to 0.
When $\lambda$ is larger than 10 $\sim$ 20, \code{level\_shift} values of 1 $\sim$ 2 or larger are needed for convergence.
The larger values for \code{level\_shift} reduce the mixing between the occupied and virtual orbits, which slows the orbital rotation with the intent of assisting convergence.
In some systems, the initial ZMP iterations will significantly perturb the potential from a trajectory towards convergence. Increasing  \code{level\_shift} to slow the orbital rotation as 
shown in Figure~\ref{fig:ZMPexample2} can minimize or remove this effect.
As \code{level\_shift} is an artificial value added to aid in convergence, its contribution is ignored when calculating properties such as the HOMO-LUMO gap and orbital energies.

\subsection{Wu-Yang method}

Performing a KS inversion with \code{kspies.wy} requires an input Mole() object and target density, as demonstrated in Figure~\ref{fig:WYexample_simple}. The potential basis (\code{pbas}) defaults to the atomic orbital basis, and \code{Sijt} is integrated analytically.
At the conclusion of Figure~\ref{fig:WYexample_simple}, the KS potential produced by maximizing $W\s$ is now accessible as an instance attribute.

\begin{figure}[!ht]
\codebox
\begin{lstlisting}
wy_a = kspies.wy.RWY(mol_1, P_tar_1)
wy_a.run()
\end{lstlisting}
\codeboxend
\caption{Minimum inputs for a WY calculation using the predefined Mole() object (\code{mol\_1}) and target density (\code{P\_tar\_1}) from Figure~\ref{fig:ZMPexample1}.}
\label{fig:WYexample_simple}
\end{figure}

\begin{figure}[!ht]
\codebox
\begin{lstlisting}
wy_b = wy.RWY(mol_1, P_tar_1, pbas='aug-cc-pV5Z')
wy_b.method = 'BFGS'
wy_b.guide = 'blyp'
wy_b.tol = 1e-7
for eta in [ 1e-3, 1e-4, 1e-5, 1e-6 ]:
  wy_b.reg = eta
  wy_b.run()
  gap = wy_b.mo_energy[5] - wy_b.mo_energy[4]
  v = wy_b.Dvb()
  print(f'eta= {eta:.1e} gap: {gap:.5f} v_grad: {v:.3f}')
\end{lstlisting}
\codeboxend
\terminalout
\begin{lstlisting}
eta= 1.0e-03 gap: 0.67420 v_grad: 1.501
eta= 1.0e-04 gap: 0.69885 v_grad: 2.741
eta= 1.0e-05 gap: 0.71409 v_grad: 7.141
eta= 1.0e-06 gap: 0.71568 v_grad: 12.703
\end{lstlisting}
\terminaloutend
\caption{Use of \code{kspies.wy} (top) and terminal outputs (bottom) using \code{mol\_1} and \code{P\_tar\_1} as created in Figure~\ref{fig:ZMPexample1}.
}
\label{fig:WYexample1}
\end{figure}

\begin{figure}[!ht]
\codebox
\begin{lstlisting}
wy_b.info()
print(len(wy_b.b))
print(wy_b.converged)
\end{lstlisting}
\codeboxend
\terminalout
\begin{lstlisting}
****Optimization Completed****
      after 325 iterations
func_value :   -128.48469083
max_grad   :      0.00000004
127
True
\end{lstlisting}
\terminaloutend
\caption{Informational calls (top) and outputs (bottom) to the WY instance from Figure \ref{fig:WYexample1}. Outputs confirms a converged KS potential.}
\label{fig:WYexample_output}
\end{figure}

A second WY example in Figure~\ref{fig:WYexample1} demonstrates additional options and uses non-equivalent basis functions.
The user specifies the potential basis set with \code{pbas}, demonstrated in the example with aug-cc-pV5Z.
The terminal output (Figure~\ref{fig:WYexample1}, bottom) shows the HOMO-LUMO gap and the smoothness of the correction potential (Eq.~\ref{eq:Reg}) for each \code{eta}.

Figure~\ref{fig:WYexample_output} calls the instance from Figure~\ref{fig:WYexample1} for information on the run, the number of basis functions used in the potential, and confirmation on convergence. The instance overwrites itself after each calculation and the reported values are for \code{eta = 1e-6}. 

Options available in \code{kspies.wy} include:
\renewcommand{\labelenumi}{(\roman{enumi})}
\begin{enumerate}
\item \code{method} (string): 
Optimization algorithm used in \code{scipy.optimize.minimize}.
Default is 'trust-exact'.
\item \code{guide} (string): 
Guiding potential. Default is 'faxc'. 
Usage is same as \code{guide} in ZMP.
\item \code{tol} (float):
Tolerance of the maximum gradient values used to determine optimization completion (Eq.~\ref{eq:WY_grad}).
Default is 1e-6.
\item \code{reg} (float):
Potential regularization weight. ($\eta$ in Eq.~\ref{eq:WY_reg})
Default is 0.
\end{enumerate}

The molecular orbital coefficients, energies, occupation numbers, density matrices, and convergence are all stored as instance attributes.
Optimized $\{ b_t \}$ values are stored as '\code{b}', and can be useful as an initial guess for subsequent calculations using different regularization strength $\eta$, as demonstrated in the for loop of Figure~\ref{fig:WYexample1}.
Other notable methods include: \code{make\_rdm1()} for generating a density matrix in the same format and method as the PySCF function of the same name, \code{info()} for printing optimization results, and \code{Dvb()} for calculating the smoothness of the correction potential (Eq.~\ref{eq:Reg}).

A WY inversion may fail if the Scipy optimizer failed to find a maximum $W\s$.
Common reasons for this failure with straightforward solutions include:
1. The given electron density is not $v\s$-representable. As the target density cannot be reconstructed with a density from single-determinant. Schipper et al. \citenum {SGB98} provide further discussion about $v\s$-representability.
2. A bad initial guess was used. By default, $\{ b_t \}$ is initialized as zero. Users can specify an initial $\{ b_t \}$ guess with the \code{b} attribute, i.e., \code{wy\_a.b = b\_init}.
For stretched molecules, an initial guess $\{ b_t \}$ from less stretched calculations may solve this failure.
3. A nearly singular Hessian was encountered. This can occur when the potential basis is very large. We recommend a gradient-based optimization algorithm for these problems, such as conjugate gradient (\code{CG}) or 
Broyden–Fletcher–Goldfarb–Shanno (\code{BFGS}).
4. An unsuitable guiding potential was used. For example, using a semi-local DFT XC guiding potential for stretched ionic bond system would lead to failure in a similar nature to how a forward KS calculation using the same functional may have trouble identifying a converged density.\cite{NSSB20}

\subsection{Regularized WY}
Regularized WY can compensate for issues resulting from unbalanced potential basis sets. \cite{HBY07}
Figure~\ref{fig:WYreg1} provides example inputs on a N$_2$ molecule, and Figure~\ref{fig:WYreg2} displays example plots used to select the regularization parameter. 
\code{kspies.wy.RWY} requires a Mole() object, a target density, and a large even-tempered Gaussian basis set for the potential calculation, all of which can be produced using PySCF.

\begin{figure}[!ht]
\codebox
\begin{lstlisting}
mol_2 = gto.M(atom='N 0 0 0 ; N 1.1 0 0', basis='cc-pVDZ')
mf = scf.RHF(mol_2).run()
P_tar_2 = mf.make_rdm1()

PBS = gto.expand_etbs([(0, 13, 2**-4 , 2),
                       (1, 3 , 2**-2 , 2)])
wy_c = kspies.wy.RWY(mol_2, P_tar_2, pbas=PBS)
wy_c.tol=1e-7

import numpy as np
etas = [ 1/2**a for a in np.arange(5,27,1)]
v = np.zeros(len(etas))
W = np.zeros(len(etas))
for i, eta in enumerate(etas):
  wy_c.reg = eta
  wy_c.run()
  v[i] = wy_c.Dvb()
  W[i] = wy_c.Ws

wy_c.reg = 0.
wy_c.run()
Ws_fin = wy_c.Ws

import matplotlib.pyplot as plt
fig, ax = plt.subplots(2)
ax[0].scatter(np.log10(Ws_fin-W), np.log10(v))
ax[1].scatter(np.log10(etas), v*etas/(Ws_fin-W))
\end{lstlisting}
\codeboxend
\caption{Regularized WY calculations require an additional potential basis set and eta parameter. Selection of eta using a trial and error process as plotted in Figure~\ref{fig:WYreg2} is recommended.}
\label{fig:WYreg1}
\end{figure}

\begin{figure}[!ht]
\includegraphics[width=1\columnwidth]{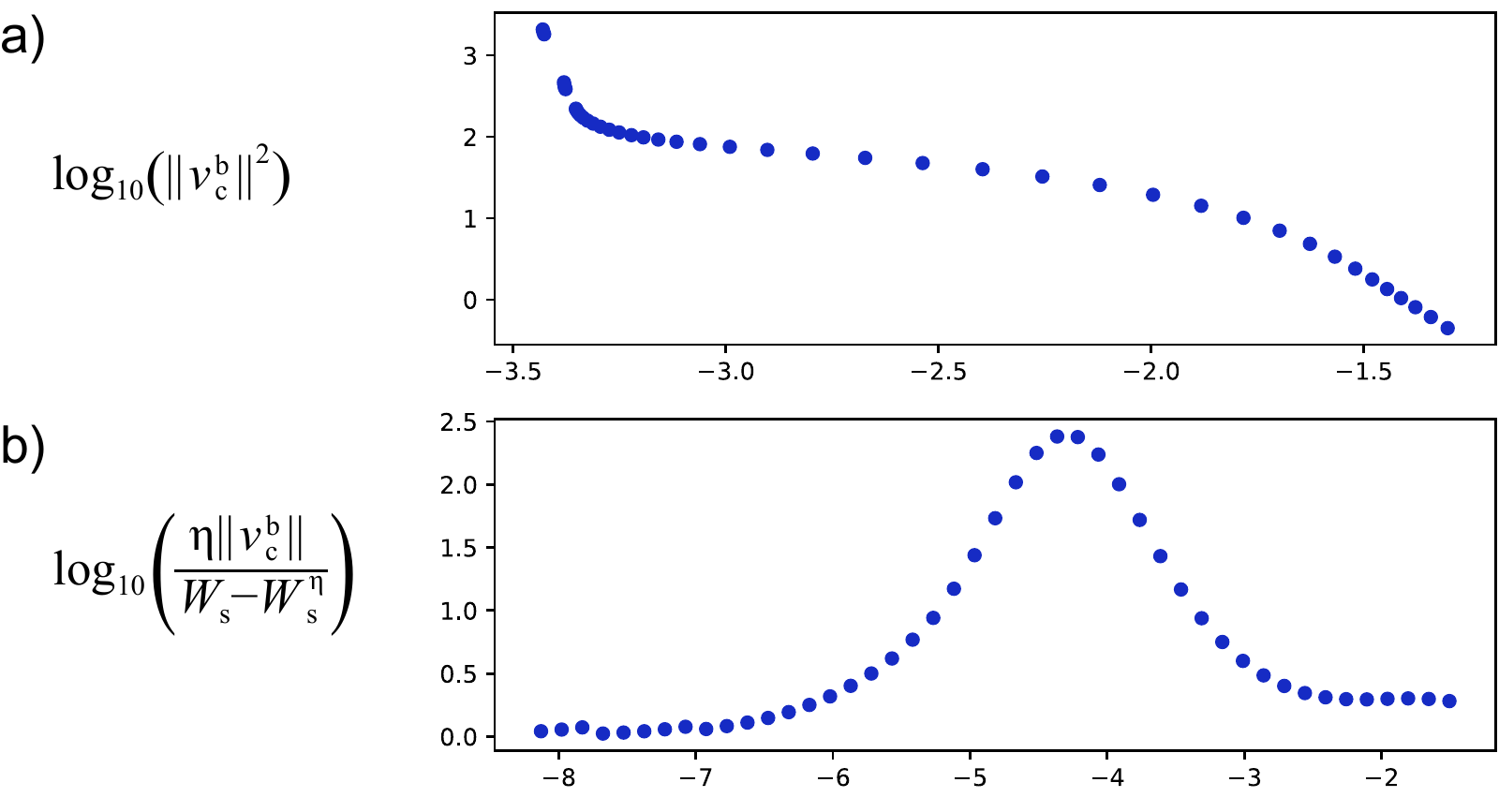}
\caption{
The L-curve (a) and its reciprocal derivative (b) used for selecting eta. The plots are outputs from Figure ~\ref{fig:WYreg1} and displayed here next to the formulas used in the plot.
The horizontal axis of (a) is log($W\s-\overline{W}\s^{\eta}$) and of (b) is log($\eta$).
The optimal $\eta$ value is the maximum of the reciprocal derivative, Eq.~\ref{eq:rec_deriv}, approximately $10^{-4.2}$ as visible in (b).
}
\label{fig:WYreg2}
\end{figure}

The optimal regularization parameter, $\eta$
can be identified using the L-curve method of,\cite{BHCY07} where the maximum value of the reciprocal derivative should be selected.
As $\eta$ decreases, $\overline{W}\s^\eta$ approaches $W\s$ and the curvature of the KS potential increases (i.e. increasing $\| \nabla v_C\b \|^2$). The need for regularization arises because an unbalanced potential basis set may cause $\| \nabla v_C\b \|^2$ to increases sharply, but is not guaranteed to correspond with a significant decreases in $W\s - \overline{W}\s^\eta$ (see Figure~\ref{fig:WYreg2}a).
This implies that the increase in potential curvature results from unphysical oscillation.
According to,\cite{HBY07} an optimal $\eta$ gives a maximum reciprocal slope of the L-curve, which has the analytical form \begin{equation}
(\frac{\partial \log ( \| \nabla v_C\b \|^2 )} {\partial \log ( W\s - \overline{W}\s^\eta ) })^{-1} = \eta \frac {  \| \nabla v_C\b \|^2 } { W\s - \overline{W}\s^\eta},
\label{eq:rec_deriv}
\end{equation}
as identified by Bulat et al.\cite{BHCY07}
Using an N$_2$ molecule with the cc-pVDZ basis and a HF target density, the reciprocal derivative of the L-curve plotted for each $\eta$ in Figure~\ref{fig:WYreg2}b), highlights the usefulness of this approach in selecting an optimal value; $\eta \approx 10^{-4.2}$ in the given example.

\subsection{WY for user-defined potential basis}

\begin{figure}[!ht]
\codebox
\begin{lstlisting}
def make_sto(zeta):
  def sto(coords):
    dist = np.sum(coords**2,axis=1)**.5
    return np.exp(-zeta*dist)
  return sto

pbas = []
for zeta in [ 0.25, 0.5, 1., 2., 4., 8. ]:
  pbas.append(make_sto(zeta))

mywy = kspies.wy.RWY(mol_1, P_tar_1, pbas=pbas)
mywy.run()
mywy.info()
\end{lstlisting}
\codeboxend
\terminalout
\begin{lstlisting}
Three-center overlap integral by numerical integration
n1  :   46
n2  :    6
time: 0.0 min
****Optimization Completed****
      after 9 iterations
func_value :   -128.48460973
max_grad   :      0.00000015
\end{lstlisting}
\terminaloutend
\caption{Informational calls (top) and outputs (bottom) to the WY instance from Figure \ref{fig:WYexample1}. Outputs confirms a converged KS potential.}
\label{fig:WYuserpbs}
\end{figure}
Although Gaussian functions are typically used to expand correction potential in WY due to their integration efficiency, this basis choice is not mandatory.
KS-pies can perform WY calculation on any user-defined potential basis.
Figure~\ref{fig:WYuserpbs} shows example of using Slater-type basis functions to expand the WY correction potential.
A user can designate \code{pbas} as a list of functions that take Bohr-unit xyz coordinates as inputs and return a value for each coordinates. 
Using this \code{pbas}, \code{kspies.wy} will then numerically calculate the three-center overlap integral and perform a WY calculation with it.
Note that numerical integration of the three-center overlap integral adds an initial computing cost which can be substantial for large systems.

\subsection{WY for user-defined systems}
\label{WY_userdefined_section}
KS-pies can calculate a WY KS potential for user-defined Hamiltonians in the restricted and unrestricted formalism, as demonstrated in Figure~\ref{fig:WY_userdefined}.
A user must provide a Mole() object (\code{mol}), target density density matrix (\code{P\_tar}), and a \code{Sijt} array to instantiate a user defined system, as well as the following instance attributes: Kinetic energy \code{T},
external potential \code{V}, overlap matrix \code{S}, kinetic energy matrix of the potential basis \code{Tp}, and a \code{None} override to prevent default use of a guiding potential.

\begin{figure}[!ht]
\codebox
\begin{lstlisting}
wy_d = kspies.wy.RWY(mol, P_tar, Sijt=Sijt)
wy_d.T = Kinetic_matrix
wy_d.Tp = Kinetic_matrix_potential_basis
wy_d.V = Potential_matrix
wy_d.S = Overlap_matrix
wy_d.guide = None
wy_d.run()
\end{lstlisting}
\codeboxend
\caption{An example input for running WY with a user-defined Hamiltonian, where the user has calculated all the necessary variables. See the online documentation for additional examples.
}

\label{fig:WY_userdefined}
\end{figure}

\subsection{Utility for cross-platform inputs}

The \code{kspies.util.wfnreader} function loads wfn file formats as generated by many quantum chemistry packages, such as Gaussian,\cite{g16} ORCA,\cite{ORCA} GAMESS,\cite{GAMESS} and Molpro,\cite{Molpro} and converts it into the PySCF format used in kspies.
Figure~\ref{code:wfnreader} shows an example usage of \code{kspies.util.wfnreader} to read wavefunction information from a \code{ccsd.wfn} file.
Note that system information such as nuclear coordinates, number of electrons, and basis sets stored in wfn format should be the same as in \code{mol}.
The \code{dm\_tar} output in the second line of Figure~\ref{code:wfnreader} is ready to serve as a target density for ZMP or WY.

\begin{figure}[!ht]
\codebox
\begin{lstlisting}
mo_coeff, mo_occ, mo_energy = kspies.util.wfnreader('ccsd.wfn', mol)
dm_tar = scf.hf.make_rdm1(mo_coeff, mo_occ)
\end{lstlisting}
\codeboxend
\caption{
Example for loading wfn file to PySCF. }
\label{code:wfnreader}
\end{figure}

Of course, users also can use PySCF default conversion tools to load densities that are generated with other platforms, which can extend the usability of KS-pies.

\subsection{Evaluation utility}
\label{Utility_section}

\begin{figure}[!ht]
\codebox
\begin{lstlisting}
coords = []
for x in np.linspace(0, 3, 1001):
    coords.append((x, 0., 0.))
coords = np.array(coords)

zmp_c = kspies.zmp.RZMP(mol_1, P_tar_1)
zmp_c.guide = 'faxc'
for l in [ 16, 128, 1024 ]:
    zmp_c.level_shift = 0.1*l
    zmp_c.zscf(l)
    dmxc = l*zmp_c.dm - (l + 1./mol_1.nelectron)*P_tar_1
    vxc = kspies.util.eval_vh(mol_1, coords, dmxc )
    plt.plot(coords[:, 0], vxc, label = r'$\lambda$='+str(l))
\end{lstlisting}
\codeboxend
\caption{An example script using \code{kspies.util.eval\_vh} for visualizing XC potential obtained with ZMP. Plotting and formatting commands, such as \code{plt.show()}, are omitted, but included in additional examples available in the online documentation. An example plot is shown in Figure~\ref{fig:vx}a.
\code{mol\_1} and \code{P\_tar\_1} are as defined in Figure~\ref{fig:ZMPexample1}.
}
\label{code:zmp_fa}
\end{figure}

\begin{figure}[!ht]
\codebox
\begin{lstlisting}
wy_e = kspies.wy.RWY(mol_1, P_tar_1, pbas='cc-pVQZ')
wy_e.method = 'BFGS'
wy_e.guide = 'pbe'
wy_e.tol = 1e-7

from pyscf import dft
ao2 = dft.numint.eval_ao(wy_e.pmol, coords)
vg = util.eval_vxc(mol_1, P_tar_1, 'pbe',
        coords, delta=1e-8)

for eta in [1e-2, 1e-3, 1e-4, 1e-5, 1e-6]:
  wy_e.reg = eta
  wy_e.run()
  vC = np.einsum('t,rt->r', wy_e.b, ao2)
  plt.plot(coords[:,0], vg+vC, label=str(eta))
\end{lstlisting}
\codeboxend
\caption{An example script using \code{eval\_vxc} for visualizing XC potential obtained with WY. The resulting plot is shown in Figure~\ref{fig:vx}b. 
\code{mol\_1} and \code{P\_tar\_1} are as defined in Figure~\ref{fig:ZMPexample1}.
}
\label{code:wy_pbe}
\end{figure}

\begin{figure}[!ht]
\centering
\includegraphics[width=1\columnwidth]{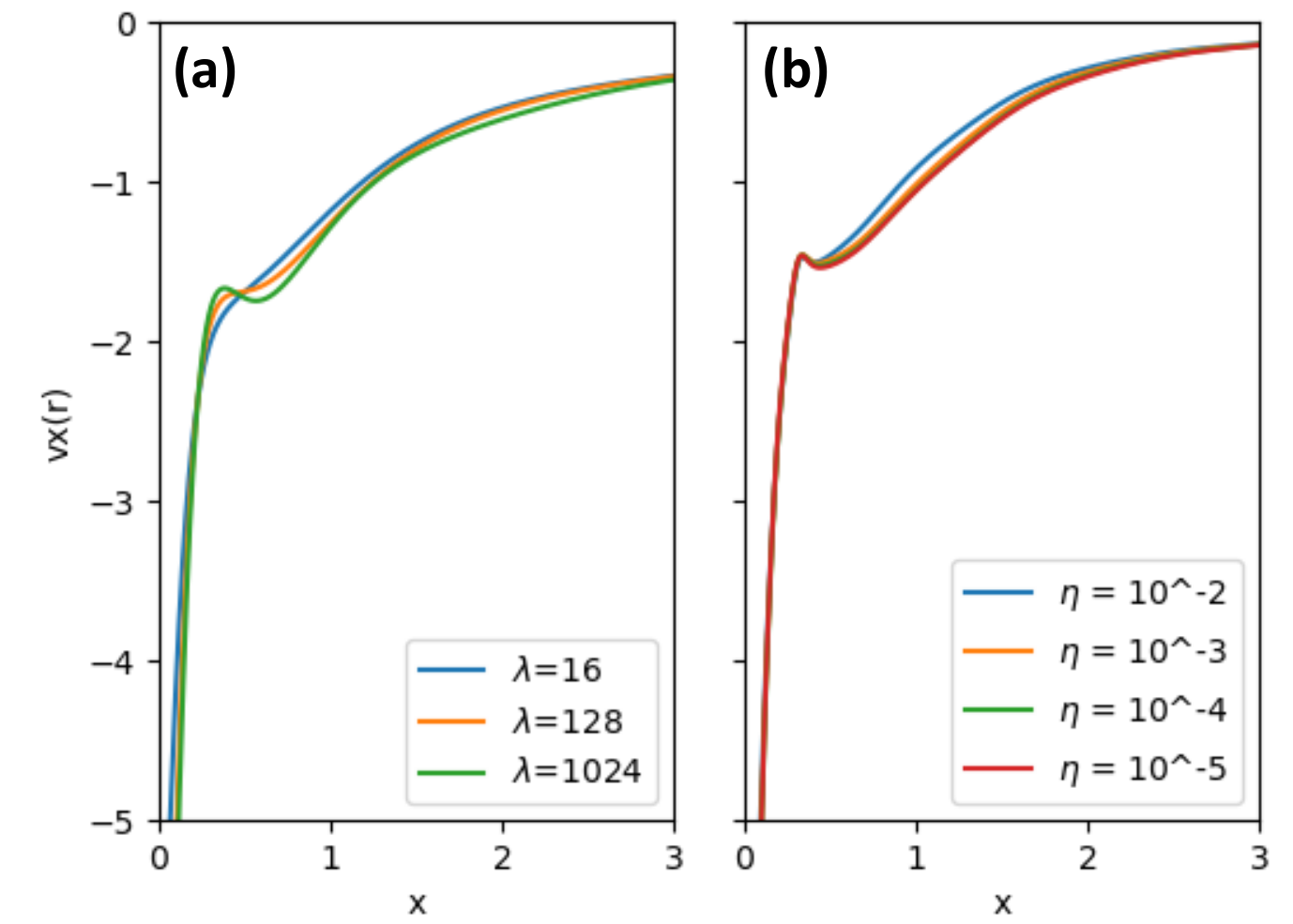}
\caption{
The exchange potential of atomic Ne produced using Figure~\ref{code:zmp_fa} (a) and Figure~\ref{code:wy_pbe} (b).
The horizontal axis denotes the distance from the Ne nucleus in Bohr, and the vertical axis denotes the exchange potential obtained from the inversion of the HF density in atomic units. These examples are included in the online documentation, and users can configure the code to display regions of interest. 
}
\label{fig:vx}
\end{figure}

The usage of \code{kspies.util.eval\_vh} function for evaluating XC potentials (Figure ~\ref{fig:vx}a) is presented in Figure~\ref{code:zmp_fa}.
Referring to the example, \code{kspies.util.eval\_vh} requires a Mole() object (\code{mol}), density matrix to calculate Hartree potential (\code{dmxc}), and specified xyz-coordinates (\code{coords}) that describe the positions of grid points that are to be calculated.
At a given $\lambda$, the ZMP XC potential can be written as
\begin{equation}
\begin{split} 
v\xc(\br) &= -\frac{1}{N}v\H[n\tar](\br)+\lambda(v\H[n^\lambda](\br)-v\H[n\tar](\br)) \\
&=v\H [ \lambda n^\lambda -  (\frac{1}{N}+\lambda)n\tar ](\br),
\end{split}
\end{equation}
indicating that the XC potential obtained from ZMP at the specific $\lambda$, is the Hartree potential of the density $\lambda n^\lambda -  ((1/N)+\lambda)n\tar$.

An example of \code{kspies.util.eval\_vxc} is used in Figure~\ref{code:wy_pbe} to create the visualization of the WY XC potential in Figure~\ref{fig:vx}b.
The finite difference required for numerical differentiation of $v_\gamma$ in Eq.~\ref{eq:vgga} and Eq.~\ref{eq:vgga_s} is set with \code{delta} (atomic units) in \code{eval\_vxc}, and defaults to $1e-7$.

Either ZMP or WY methods can be used with \code{kspies.utils.eval\_vh} and \code{kspies.utils.eval\_vxc}. Some useful possibilities beyond our current examples include using ZMP with PBE XC guiding potential, to draw PBE XC potential with \code{kspies.utils.eval\_vxc} and ZMP correction potential with \code{kspies.utils.eval\_vh}. Visualization with  \code{kspies.utils} is not limited to XC potential obtained from KS inversion, but can be used independently with KS inversion. 

%\nocite{*}
\bibliographystyle{apsrev4-2}
%\bibliography{refs}% Produces the bibliography via BibTeX.
%apsrev4-2.bst 2019-01-14 (MD) hand-edited version of apsrev4-1.bst
%Control: key (0)
%Control: author (72) initials jnrlst
%Control: editor formatted (1) identically to author
%Control: production of article title (-1) disabled
%Control: page (0) single
%Control: year (1) truncated
%Control: production of eprint (0) enabled
%

\end{document}